\documentclass[aps,pre,reprint,twocolumn,showpacs]{revtex4}

\usepackage{amsmath,amssymb,amsfonts}

\usepackage{inputenc}

\usepackage{graphicx}

\begin{document}
\title{Predicted and Verified Deviations from Zipf's law in Ecology of Competing Products}
\author{Ryohei Hisano$^{a}$, Didier Sornette$^{a}$, Takayuki Mizuno$^{b}$}%
\address{$^{a}$ETH Zurich, Department of Management, Technology and Economics, Kreuzplatz 5, CH-8092 Zurich, Switzerland 
\linebreak $^{b}$Department of Computer Science, Graduate school of SIE, University of Tsukuba, Ibaraki, Japan 
}

%\author{Ryohei Hisano and Didier Sornette}
%\affiliation{ETH Zurich,
% Department of Management, Technology and Economics, Kreuzplatz 5, 
%CH-8092 Zurich, Switzerland}

\date{July 18 2011}

%\maketitle
%\begin{article}
\begin{abstract}
Zipf's power-law distribution is a generic empirical statistical regularity found in many complex systems.  However, rather than universality with a single power-law exponent (equal to $1$ for Zipf's law), there are many reported deviations that remain unexplained.  A recently developed theory finds that the interplay between (i) one of the most universal ingredients, namely stochastic proportional growth, and (ii) birth and death processes, leads to a generic power-law distribution with an exponent that depends on the characteristics of each ingredient. Here, we report the first complete empirical test of the theory and its application, based on the empirical analysis of the dynamics of market shares in the product market.  We estimate directly the average growth rate of market shares and its standard deviation, the birth rates and the ``death" (hazard) rate of products.  We find that temporal variations and product differences of the observed power-law exponents can be fully captured by the theory with no adjustable parameters.  Our results can be generalized to many systems for which the statistical
properties revealed by power law exponents are directly linked to the underlying generating mechanism.
\end{abstract}

\maketitle

Power-law distributions constitute ubiquitous statistical features of many natural and social complex phenomena [1-3].  The probability distribution function $p(s)$ of a random variable, 
\begin{equation}
p(s) \sim 1 / s^{1+\mu}~,
\end{equation} 
describes a particularly slow decay with $s$ of the probability $p(s)ds$ that the random variable is found in one realization between $s$ and $s+ds$.  A power-law distribution is such that any of its moment of order $q$ larger than the power-law exponent $\mu$ is mathematically infinite.  Among power-law distributions, Zipf's law, corresponding to $\mu=1$, has been proposed as a fundamental characteristic for many systems [4-6].  Zipf's law implies that we no longer have finite means in infinite systems, or that the means are strongly system size dependent in the real world.

Motivated by its apparent ubiquity and interesting features, many efforts have been made to attempt explaining the existence of power-law distributions.  One of the general mechanisms to generate power-law distributions is embodied in the multiplicative stochastic growth models, having Gibrat's rule of proportional growth [7]
as a key ingredient.  Expressed in continuous time, Gibrat's rule is equivalent to the well known geometric Brownian motion
\begin{equation}
dS(t) = S(t) \left( r \, dt + \sigma \,  dW(t)\right)~,
\label{wbtwr}
\end{equation} 
where $r$ denotes a drift, $\sigma$ is the volatility (or standard deviation) and $W(t)$ is a standard Wiener process.  However, Gibrat's rule alone cannot generate a stable power-law distribution, since the solution of the process (\ref{wbtwr}) leads to a non-stationary log-normal distribution.

Since Herbert Simon's work  [8-10] that extended previous attempts at explaining Zipf's law [4,11-13], a huge literature followed that is spread across a variety of disciplines [14-23]. This has led to the understanding that an apparently minor modification in the multiplicative process applying just when $S(t)$ becomes small suffices to ensure a stationary power-law distribution.  One simple way to ensure this condition is by adding an additive noise in the system of recursive equations which leads to the so called Kesten process [23-25].  This process has been studied extensively in the physics literature because some nonlinear dynamic systems can be approximated by such equations.  We now know that under very general conditions concerning the nature of the perturbation for small $S(t)$'s and on the distribution of the stochastic growth factors in the multiplicative process, the power-law exponent could be obtained as a function of the distribution of the multiplicative growth factors [26-28].

However, it should be realized that the derivation of a power-law distribution based on expression (\ref{wbtwr}) augmented by some modifications for small $s(t)$'s, such as in the Kesten process [23], relies crucially on a view of dynamics in which all entities are born at the same instant [29]. 
Indeed, the distribution of sizes calculated by this kind of approach is nothing but the distribution of the sizes that a single entity explores
over time, as a result of the stochastic growth mechanism. Since, in empirical data, the distribution of sizes is a cross-sectional
snapshot of the set of sizes of an ensemble of entities at a given time, the correspondence between the two is based on 
an argument of ergodicity, that is, the distribution of $S(t)$'s for a single realization along time is the same as that obtained from a snapshot at a given time for an ensemble of entities that are supposed to be statistically equivalent.  
But this cannot be true in general if entities are born at different 
times, and thus have different ages and different average sizes. In the presence of a flux of entrant entities,
the population at a given time is a mixture of just-born, young, mature and old entities. Previous attempts to explain
the occurrence of power laws and of Zipf's law from Gibrat's law of proportional growth (plus some ingredient for small sizes)
fail to account for the unavoidable and significant fact that entities that grow are continuously born and end up dying.
The condition of simultaneous births is clearly counterfactual for cities, firms, species, and many other examples for which power-laws and zipf's law are observed.

To account for the fact that entities of any kind are born and die, Malevergne et al. [30], extending the classical theory of H. Simon, proposed a general framework where (i) there exists a random birth flow of entities, (ii) entities grow according to equation (2), and (iii) entities exit the system according to possible different mechanisms, including a minimum size threshold as well as exogenous shocks controlled by some stochastic process with hazard rate $h$.  This theory finds that Zipf's law can be seen as either 1) the statistical property corresponding to the condition when stochastic growth dominates over other factors or 2) the statistical property corresponding to the condition of stationary growth of the system (for more detail see model section).  Malevergne et al. [30] also provided the explicit functional relationship between the power-law exponent and its key parameters, predicting deviations from Zipf's law due to various sources (see equation (\ref{trj4jj}) below).  

The main motivation of Malevergne et al. [30] was to introduce a more realistic reduced-form model of firm growth
to predict the firm size distribution within a generally growing economy. The main result was the discovery that Zipf's law
corresponds to the delicate balance in which the economy grows at its maximally sustainable growth rate and that
deviations from Zipf's law can be quantitatively explained from the parameters encoding the growth, birth and death 
properties of the firms. The goal of the present paper is to provide what we believe is the first empirical test 
of this model, using a data set collected from a very different setting. In this way, we further demonstrate the generality
of the predictions of Malevergne et al. [30], which should apply to any system in which birth, death and stochastic growth occur.

To test and find an application of this theory, it is crucial to find a data set in which all the relevant process (birth flow, death flow and stochastic growth) could be measured explicitly.  Here, we use a unique data set from the Japanese consumer electronics market containing all the information needed to test the model.  Hisano and Mizuno [31] have recently reported that product market shares obey power-law distributions that possess an interesting time evolving nature.  However, such time varying nature of the market share distributions and how it could be understood from the underlying process has not been fully explored.  Using this data set, we first verify tha
t the ingredients of the model are indeed present.  We then show that the theoretical prediction (\ref{trj4jj}) is in excellent agreement with the empirical power-law exponents found in the tail of the market share distribution of different electronic products with no adjustable parameters.  It also accounts very well for their variations with time giving deeper insights into the nature of how the empirical power-law distributions are formed and evolves with time in this system.  We also identify the cause of the deviation from a genuine power-law distribution, which gives insights into the importance of flow of births, which further supports the theory [30] in contrast with alternative approaches [23,26,27] which neglects the birth and death processes.

\section{Model}

Consider a population of entities (firms, cities, projects, products and so on), which can take different forms and can be applied in many different contexts.   The theory is based on the following assumptions [29].  
\begin{enumerate}
\item[A 1] Gibrat's rule of proportional growth holds.  This implies that, in the continuous time limit, the size $S_i(t)$ of the $i$th entity at time $t$, conditional on its initial size, is solution to the stochastic differential equation (2), where drift ($r$) and volatility ($\sigma$) are the same for all entities, but the Wiener process $W_i(t)$ is specific to each entity.

\item[A 2] Entities are born with initial sizes $s_{0,i}$ that are independent and identically distributed random variables.  It could be shown that, to a large extent, the characteristics of the distribution of initial sizes is irrelevant to the shape of the upper tail of the steady-state distribution of entities [30]. 

\item[A 3] There exists a random birth flow of entities following a Poisson process (extensions to a vast class of non-Poisson processes do not alter key results [29]).
\item[A 4] Entities exit at random with a constant hazard rate $h$, which is independent of the size of the entities.
\end{enumerate}

Under these conditions, we could prove that for times larger than
\begin{equation}
t_{\rm transient} = \left[ \left(r - \frac{\sigma^2}{2} \right)^2 + 2 \sigma^2 h\right]^{-1/2}~,
\end{equation}
the size distribution of entities follows an asymptotic power-law with tail index $\mu(TH)$ 
\begin{equation}
\mu(TH) := \frac{1}{2} \left[\left(1 - 2 \cdot \frac{r}{\sigma^2}\right) + 
\sqrt{\left(1 - 2 \cdot \frac{r}{\sigma^2}\right)^2 + 8 \cdot \frac{h}{\sigma^2}} \right]~.
\label{trj4jj}
\end{equation}

There are two possible ways to obtain Zipf's law.  One way is when $\sigma$ is large compared to both $|r|$ and $h$.  When $\sigma$ is large, then irrespective of all other parameters, the power-law exponent converges to $1$ as $\sigma \to \infty$ either from above or below depending on the other parameters.  Under this condition, Zipf's law can be seen as a result of large stochastic multiplicative excursions that dominate the dynamics as already reported in [6].   Deviations from Zipf's law implies that the stochastic growth component is not large enough.  In the context of time evolving power-law distributions, and assuming that the hazard rate is constant over time,  the increasing or decreasing trend of the power-law exponent away from Zipf's law value could be explained by the evolution of the parameters $r$ and $\sigma$ that
control the stochastic growth process (as summarized in equation (2)). Thus, the evolution of 
the exponent can provide fruitful insights into how the growth dynamics itself changes with time, as
we shall see below.

The second way to obtain Zipf's law is when the equality $r=h$ holds.  This condition, as discussed in [30], could be understood as a balance condition which ensures the stationary growth of the system.  In order to understand how Zipf's law
is related to this equation, we
notice that $r-h$ represents the average growth rate of incumbent entities.  Indeed, considering an entity present at time $t$, during the next instant $dt$, it will either exit with probability $h\cdot dt$ (and therefore its size declines by a factor $-100\%$) or grow at an average rate equal to $r \cdot dt$, with probability ($1 - h \cdot dt$).  The coefficient $r$ is therefore the conditional growth rate of entities, conditioned on not having died yet.   Then, the unconditional expected growth rate over the small time increment $dt$ of an incumbent entities is $(r -h) \cdot dt+O(dt^2)$.  Hence the statistically stationary regime, in the presence of a stationary population of entities, corresponds to condition $r = h$ (for a more general and detailed illustration of this condition, we refer to [30]).  Although this condition would probably never hold for product markets (and average growth rate of  ``market share" will not be for long bigger than its hazard rate), this viewpoint is also interesting in a wide range of applications since, in this framework, the power-law exponent can be seen as a remarkable statistical signature providing information about the stationary or non-stationary growth of the corresponding system.

Our strategy is to find an empirical dataset in which all ingredients of the theory can be verified and measured explicitly.  Our dataset from the Japanese consumer electronics market contains all necessary ingredients.  It consists of scanner data from 23 different consumer electronics chains in Japan collected by a private company, BCN Inc.  This data set covers about 45\% of all consumer electronics chains in Japan including over 1400 retail stores.  The data were recorded daily between October 1, 2004 and April 30, 2008.  The dataset provides time series of total sales per retail store for every product sold, including manufacture and models.  We believe this data set is sufficient to represent the true dynamics of Japanese consumer electronics market.

\section{Validation of the model assumptions}

\subsection{Empirical power laws}

Fig. 1 shows the cumulative distribution of market share $S$ of digital camera for the top-selling models observed on a daily scale.  Market share was defined by the daily sales volume of a generic product such as  ``digital camera", ``mouse", and so on.  Note also that to keep consistency with Hisano and Mizuno [31], we only analyzed products which are ``alive" by the definition described below when analyzing market share distributions.  Products shown in Fig. 1 account for roughly 75-85\% of the total sales making it natural to focus on these top-selling products.  Each date plotted in Fig. 1 was chosen, following [31], from the periods when the power-law behaviors describing the tail of the distributions observed on a daily scale are stable.  As written in the same paper, the tails of market share distributions sometimes deviate from the pure power-law form, but the power-law behavior repeatedly reappears, proving that it is insufficient to characterize market share distributions with a simple lognormal distribution (which has qualitatively different tail characteristics, called for instance ``thin-tailed'' in the
mathematical literature as opposed to the ``heavy-tailed'' power laws).

\begin{figure}[!htb]
\begin{minipage}{0.45\hsize}
\begin{center}
\includegraphics[width=\hsize,bb= 14 6 150 128]{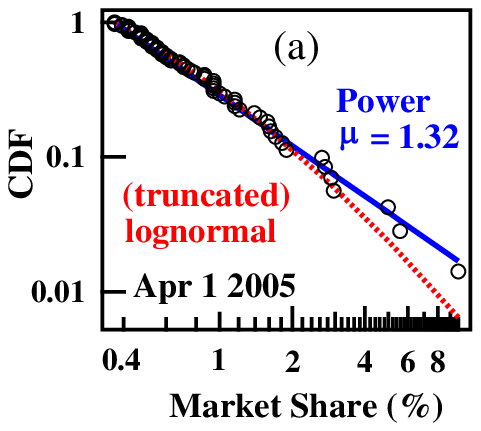}
%+17
\end{center}
\end{minipage}
\begin{minipage}{0.45\hsize}
\begin{center}
\includegraphics[width=\hsize,bb=  11 5 147 127]{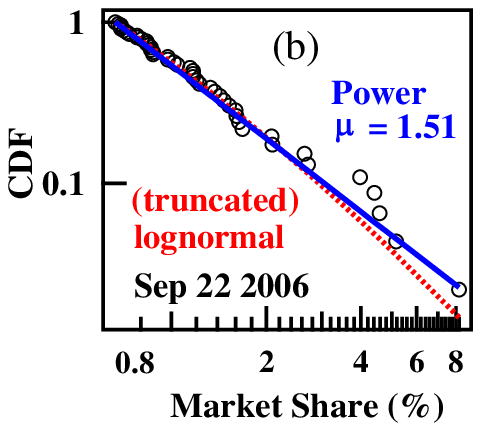}
%+17
\end{center}
\end{minipage}
\begin{minipage}{0.45\hsize}
\begin{center}
\includegraphics[width=\hsize,bb=   9 6 145 128]{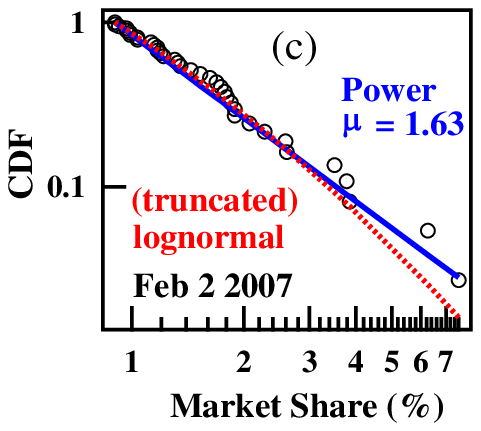}
%+17
\end{center}
\end{minipage}
\begin{minipage}{0.45\hsize}
\begin{center}
\includegraphics[width=\hsize,bb= 14 6 150 128]{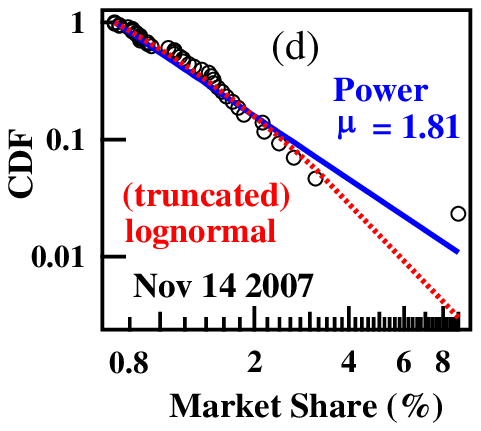}
%+17
\end{center}

\end{minipage}
\caption{(color online) Complementary cumulative distribution functions  (CDF) of market share of digital cameras measured for four snapshots on (a) April 1 2005, (b) September 22 2006, (c) February 2 2007 and (d) November 14 2007.  Maximum likelihood estimates of a power-law distribution are shown by the blue (continous, upper) lines and the maximum likelihood estimates of a truncated (at the left) lognormal distribution are shown by the red (dashed, bottom) lines.  In all cases, the empirical distributions go beyond the exploration of a truncated lognormal.}
\end{figure}

\subsection{Verification of assumption 1}

To test our model, we first verify Eq. 2, which suggests that, for sufficiently small time intervals $\Delta t$, the mean $<\Delta S>$ and the standard deviation of $\Delta S$ of the increment of the size $S$ (market share) are both proportional to $S$.  Fig. 2 plots the average and standard deviation of $\Delta S$ as a function of $S$, setting $\Delta t$ equal to one week, confirming that Eq. 2 holds. We note that the slight curvature observed on the left panel 
is really of no consequence as it can be shown [29] that 
the power law tail is controlled by the proportional growth of the stochastic component, i.e.,  the standard deviation of $\Delta S$, when the latter has a dominant behavior as is the case here since $<\Delta S>$ is decreasing with  $\Delta t$ 
while $\Delta S$ is an increasing function.

\begin{figure}
\begin{minipage}{0.45\hsize}
\begin{center}
\includegraphics[width=\hsize,bb= 15 8 157 127]{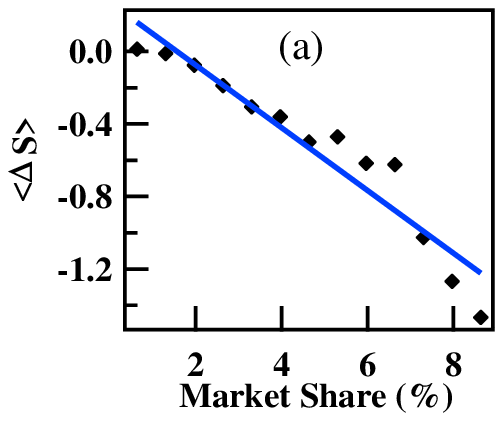}
%+17
\end{center}
\end{minipage}
\begin{minipage}{0.45\hsize}
\begin{center}
\includegraphics[width=\hsize,bb=  14 8 152 127]{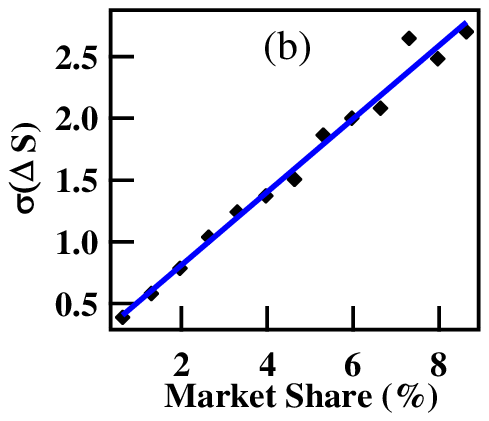}
%+17
\end{center}
\end{minipage}
\caption{(color online) Test of Gibrat's law of proportional growth for market share $S$ of products until January 31 2008.  The left panel (i.e. (a)) shows the test for the mean of the the increments (i.e. $<\Delta S>$) versus its current size.  The right panel (i.e. (b)) depicts the test for the stadnard deviation of the increments (i.e. $\Delta S$) versus its current size.  In both panels the blue (continuous) line show the OLS fit to the data points.  $R^{2} $ values are 0.913 for the left panel (i.e. (a)) and 0.984 for the right panel (i.e. (b)).}
\end{figure}

\subsection{Verification of assumption 2}

Next, we verify that the stochastic growth is indeed the determinant of the empirical power-law distribution, and
not some initial build-in effect or distribution.  Market share of products at birth varies among products due to their past brand images.  Thus, it is important to verify that the empirical distribution observed in Fig. 1 is a result of the stochastic multiplicative growth process and not of the distribution of its initial market share at birth.  To verify this, Fig. 3 shows the empirical distribution of market share of newly born products, taking market share of the first Sunday two weeks after its birth as a proxy.  Compared to the distributions found in Fig. 1, the distribution depicted in Fig. 3 has a much thinner tail, 
even thinner than log-normal, confirming that the distribution of future market share does not result from the distribution of market share of products at birth.  We hypothesize that this implies that the stochastic growth process causes the power-law behavior that appear as the market shares of the products evolve in time.

\begin{figure}[!htb]
\begin{minipage}{\hsize}
\begin{center}
\includegraphics[width=0.5\hsize,bb=  17 5 157 127]{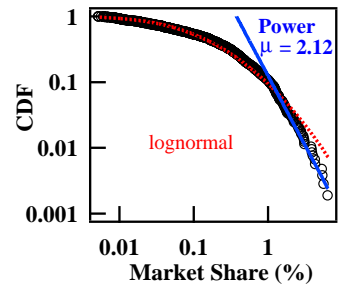}
%+22
\end{center}
\end{minipage}
\caption{(color online) Distribution of initial market product shares, taking the market share value of the first Sunday occurring two weeks after the birth of a given product as a proxy.  The distribution of initial market share distribution is shown by the black circles.  Blue (continuos) line shows maximum likelihood estimate of a power-law distribution and the red (dashed) line shows maximum likelihood of a lognormal distribution.}
\end{figure}

\subsection{Verification of assumption 3}

Fig. 4 shows three and six month moving average values of birth flows of digital cameras observed on a weekly time scale.  It displays a periodic behavior with peaks in February and August, just before the selling seasons starts.  The birth date of a product is defined by the first date when we can detect it in our data set.  To verify assumption 3, we test whether the birth flow follows an inhomogeneous Poisson process with rates varying only in a periodic manner.  This could be performed by a time rescaling method which transforms an inhomogeneous Poison process into a homogeneous Poisson process with unit rate [32].  Here, we chose the function $\lambda(t)=a(1+b \sin(ct))$ as describing the time varying rates.  The comparison between the fitted curve and moving average are shown in Fig. 4.  The Kolmogorov-Smirnov statistic of the statistical test that the waiting time of the rescaled process obeys an exponential distribution with rate 1 is 0.082, which is slightly above the critical value ($\alpha_{0.01}=0.051$) implying that the null hypothesis is only slightly rejected.  Moreover, measuring the growth 
of the Poisson rate intensity by 
the parameter $d$ in the expression of the time evolution of the Poisson intensity $\lambda(t)=v_{0} \exp(dt)$, which Malevergne et al. consider in their theory [30], we find an estimate of the parameter $d$ about $0.001$, which is statistically indistinguishable from $0$.   Similar conclusions holds for other products as well.

\begin{figure}
\begin{minipage}{\hsize}
\begin{center}
\includegraphics[width=0.75\hsize,bb= 15 19 297 157]{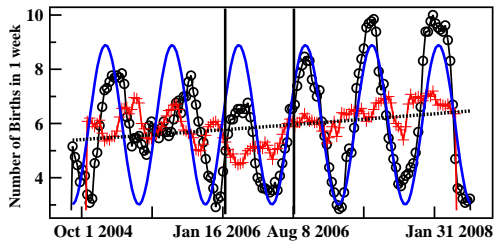}
%+22
\end{center}

%\label{fig::oneichi}
\end{minipage}
\caption{(color online) Birth flows measured at the time scale of one week.  The black circles denote 3 month moving average and the red cross denote 6 month moving average.  The blue (continuous) line shows the fitted curve $\lambda(t)=5.95*(1-0.49*\sin(0.23*t))$ and the black (dashed) curve corresponds to $\lambda(t)=5.34*\exp(0.001*t)$.  Since $2*\pi/0.23 = 27.3 (weeks) =191.13 (days)$, we can see that the periodicity is approximately 6 month.  Notice also the sharp drop of birth flows found during January 16 2006 to August 8 2006.}
\end{figure}

\subsection{About assumption 4}
Products eventually get old, lose their competitiveness, and leave the market.  It is clear from our daily experiences that this turnover speed varies among product markets, making it one of the key parameters characterizing this system.  Here, we define the death of a product as the date when its cumulative sales reaches 90\% of its total sales in the whole data set.  To avoid incorrectly identifying deaths that appear due simply to the end of our dataset, we exclude from the lifetime analysis all products that died in the last four months of our dataset and we characterize turnover speed by analyzing the distribution of product lifetimes.  

Identifying the timing of death for the entire product set is complicated due to the large statistical fluctuations found for the lower-selling products.  While the total number of products sold in a day fluctuates widely from 3,500 to 10,000 (for digital cameras) the sales volume for the lower-selling products varies from 0 to 15 sales per day, making it extremely difficult to estimate the correct timing of death.  Moreover, lower-selling products also include the extremely expensive ones (which are seldom sold), all kinds of limited editions (i.e. tourist edition, foreign use edition, limited color edition, anniversary edition, etc.) and tie-in sales products (i.e. camera with printer kit or zoom lens, and so on), which we want to exclude in our analysis because we are focusing on the asymptotic behavior of the distribution (i.e. dynamics of the top-selling products).  In order to overcome this problem and estimate the effective hazard rate which affects the asymptotic behavior of the distribution, we separate the top-selling products from the lower-selling products.  Thus in this paper, we only consider products that attained an overall market share of 0.5\% or higher during their lifetime (this could be calculated by taking the average market share a product attained during its lifetime), when we analyze the lifetime distribution of products.  Fig. 5 shows two scatter plots describing 1) the relationship between product lifetime and overall market share (as defined above) and 2) the relationship between product lifetime and the market share of products at the time of death, for all the digital cameras found in our dataset.  We can see that, for the products that attained more than 0.5\% of the market share, the lifetime of a product seems to be independent of the size, making assumption 4 reasonable at least for the top-selling products.  Of course, this does not validate assumption 4 for the entire product set because, after all, we are separating the top-selling products from the lower-selling ones, for the reason mentioned above. But Fig. 5 does show that assumption 4 is fair enough for the top-selling products that we are focusing on (i.e. corresponding to the asymptotic behavior of the distribution).  Hence we can assume that assumption 4 holds as well.

The timing of product exits is also complicated by the presence of periodicity.  In addition, there are tactical adjustment of product pricing, in response to market share dynamics.  In particular, a product suffering from a unlucky sequence of low sales may see its wholesale price lowered by its provider, thus making it easier for retail stores to post lower price, in order to clear the stocks and also probably not to lose presence in the market.  A typical empirical example of this kind of dynamics is shown in Fig. 6, with the time evolution of the average posted price of a specific digital camera.  We can see that, when the market share of this product decreased to around 2\%, the average posted price started to fall sharply, thus preventing the market share of that product from vanishing.  This in turn resulted in prolonging its lifetime.  Due to this effect, the lifetime distribution of products behaves differently at the lower end of its distribution, but we find that asymptotically it is close to an exponential distribution.  Hence, we focus on the asymptotic behavior of this distribution and estimate the hazard rate using a fit with an exponential distribution calibrated over a time interval no longer than 38 weeks (see left panel of Fig. 6).  Note also that the empirical distribution of the lifetimes of digital cameras suggests a stable trend during the whole data set, which will become relevant later on.

\begin{figure}[!h]
\begin{minipage}{0.49\hsize}
\begin{center}
\includegraphics[width=\hsize,bb= 27 12 326 219]{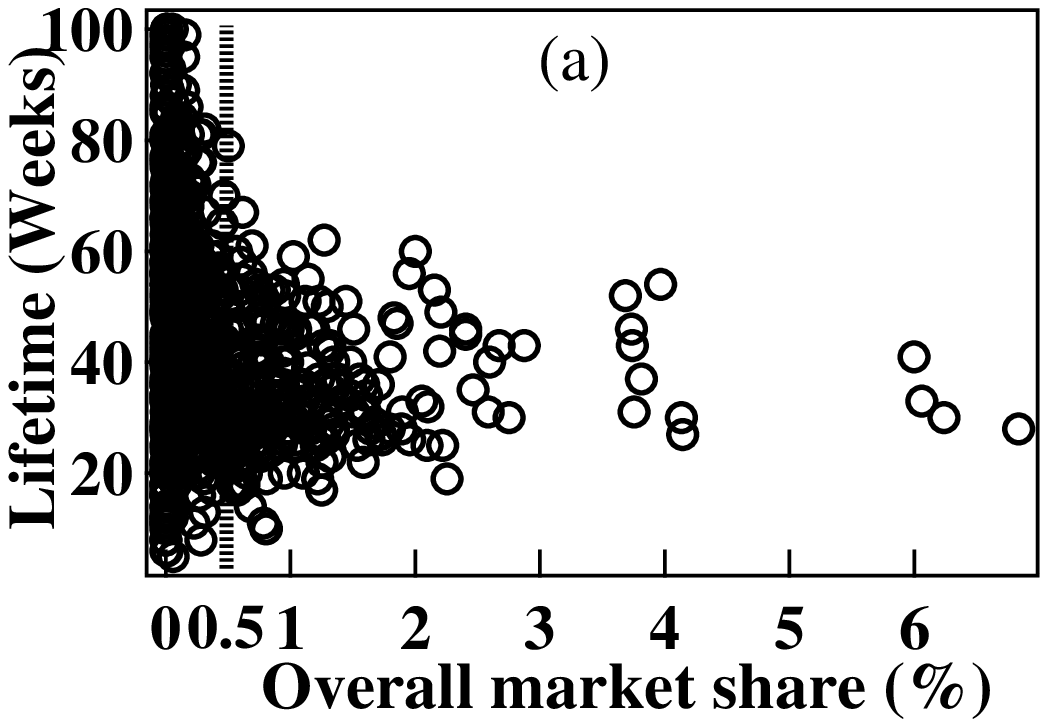}
\end{center}
\end{minipage}
\begin{minipage}{0.49\hsize}
\begin{center}
\includegraphics[width=\hsize,bb=26 10 357 223]{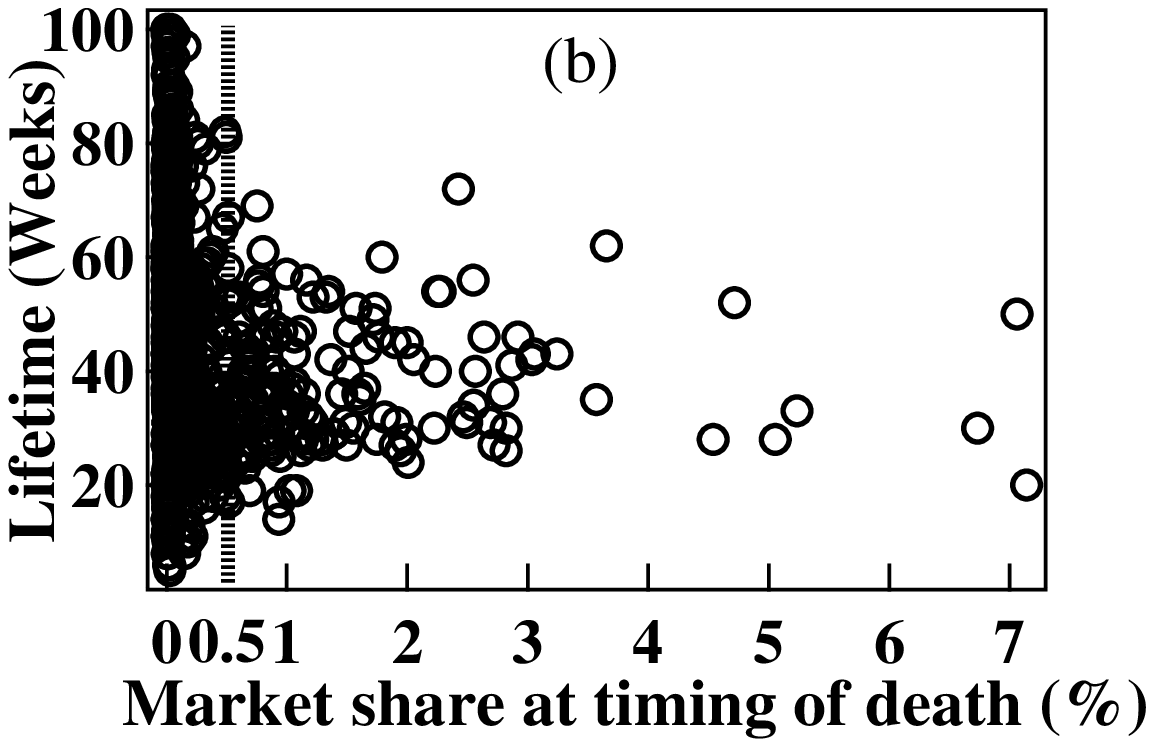}
\end{center}
%+0

\end{minipage}
\caption{Scatter plots describing (a) the relationship between product lifetime and overall market share (as defined in the manuscript) and (b) the relationship between product lifetime and the market share of products at the time of death for all the digital cameras that died during the period January 2005 to December 2007.  Note that, for products that attained more than 0.5\% of the market share, size independence of lifetime seems to hold.}
\end{figure}

\begin{figure}[!h]
\begin{minipage}{0.45\hsize}
\begin{center}
\includegraphics[width=\hsize,bb= 15 5 154 129]{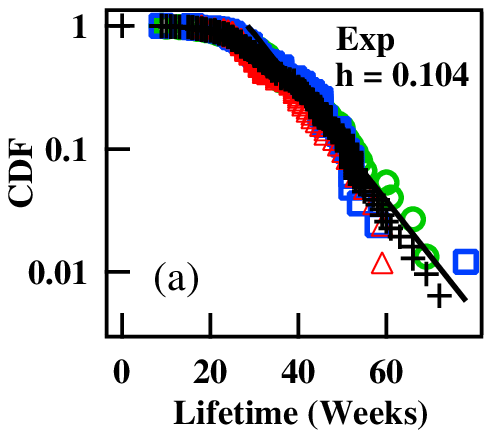}
\end{center}
\end{minipage}
\begin{minipage}{0.5\hsize}
\begin{center}
\includegraphics[width=\hsize,bb=17 26 187 152]{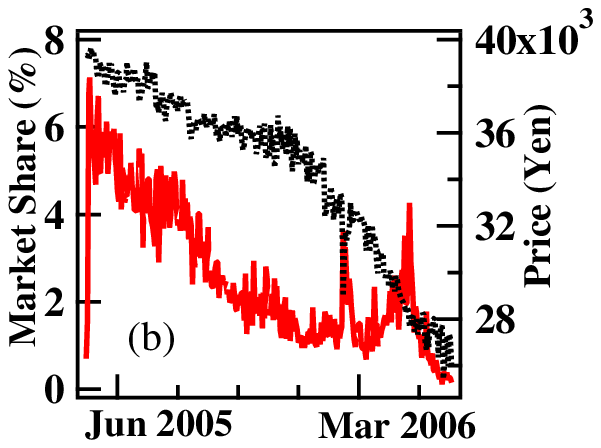}
\end{center}
%+0

\end{minipage}
\caption{(color online) The left panel (i.e. (a)) depicts the complementary cumulative distribution function (CDF) of lifetimes of digital cameras using the definition mentioned in the text.  Green circles denote the distribution of lifetimes of products which died during 2005, blue squares correspond to 2006, red triangles to 2007, and black crosses mixing all of these years together.  The right panel (i.e. (b)) depicts the time evolution of the market share and of the average posted price movement for a particular digital camera.  The (continuous) red line depicts the market share (left vertical axis) and the black (dotted) line shows the average posted price over all retail stores (right vertical axis).}
\end{figure}

\section{Tests of the theoretical prediction (\ref{trj4jj})}

\subsection{Direct test}

We now test whether the model can predict correctly the power-law exponents of the distributions shown in Fig. 1.  The drift ($r$) and volatility ($\sigma$) were estimated as the mean and standard deviation of the set of weekly growth rates using 6 month time windows, for products with market shares larger than 0.5\% (this is roughly where the power-law regime starts).  The parameter $h$ was estimated using a 1 year time window and estimating the lifetime distribution of products which died during that time window as explained above.  Using the empirically determined values of $r$, $\sigma$ and $h$, we can now test the theoretical prediction (\ref{trj4jj}).  Table 1 compares the maximum likelihood estimate obtained from the empirical distribution and the theoretically predicted power-law exponent  (\ref{trj4jj}).  It shows excellent agreement.  From this result, we conclude that the model is able to explain quantitatively the power-law exponent found in market share distribution of products.

\begin{table}[ht]
\caption{Comparison between the theoretical predicted power law exponents $\mu(TH)$ and the empirical exponents $\mu(MLE)$.  
The later exponents are obtained by direct estimation of the empirical distributions.
 Numbers in brackets are the 95\% confidence interval estimated by bootstrapping.
 The theoretical values $\mu(TH)$ are determined by reporting the independently found values of $r$, $\sigma$ 
 and $h$ in the theoretical prediction given by equation (4).}
\centering
\begin{tabular}{|c|p{1.7cm}|p{1.7cm}|p{1.7cm}|p{1.7cm}|}
\hline
Date & Apr 1 2005 &Sep 22 2006 & Feb 2 2007 &Nov 14 2007\\
\hline
$r$ & -.01[-.024,-.005] &-.003[-.014,.01]&-.0017[-.011,.008]&-.022[-.029,-.012] \\
\hline
$\sigma$ & 
.66[.592,   .728] & .561[.487,   .634]& .477[.427,   .535]& .447[.384,   .531] 
 \\
\hline
$h$ & .0871[.069,    .116]&.11[.083,    .1474]&.116[.088,   .156]&.119[.0905,   .172] \\
\hline
$\mu(MLE)$&1.35($\pm 0.1$)&1.5($\pm 0.1$)&1.6($\pm 0.1$)&1.8($\pm 0.1$)
 \\

\hline
$\mu$ (TH)) &1.342[1.254,   1.463]&1.489[1.356,   1.668]&1.638[1.483,    1.846]&1.863[1.624,   2.166]\\

\hline
\end{tabular}
\label{table:predictedmu}
\end{table}

\subsection{Variation of the power-law exponent}

Table 1 suggests that the drift $r$ and volatility $\sigma$ play crucial roles in creating the variation of the power-law exponent.  To investigate this further, we explore how the predicted power-law exponent behaves dynamically with time by adding three more products to our analysis (liquid crystal type TV, mouse and keyboard).  For each product and for each date, we open a 6 month and 1 year time window and estimate the time evolution of the average growth rate ($r(t)$) and its standard deviation ($\sigma(t)$) accordingly. Fig. 7 depicts a scatter plot showing the relationship between drift ($r(t)$) and volatility ($\sigma(t)$) for the four products mentioned above.  We confirm that the two parameters vary with product sets and also with time.

\begin{figure}
\begin{minipage}{\hsize}
\begin{center}
\includegraphics[width=0.55\hsize,bb=  19 9 332 276]{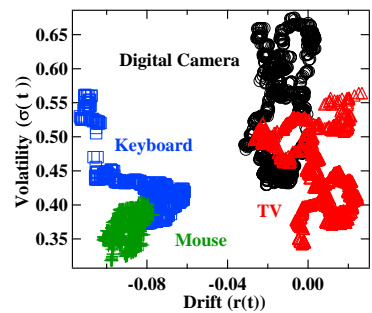}
%+22
\end{center}

%\label{fig::oneichi}
\end{minipage}
\caption{(color online) Scatter plot showing the relationship between drift ($r$) and volatility ($\sigma$) for the period November 2004 to December 2007.  Black circles denote digital cameras, red triangles represent TV's, blue squares code for keyboards, and green cross for mouse products.  The drift and volatility for digital cameras, mouse products and TV's were calculated by opening a 6 month time window and conditioning on when market share is above 0.5\%.  For keyboards, this level was slightly increased to 0.75\% because, while the mode number of products sold each day during 2007 is around 3500-5000 for other products, it was only 1300 for keyboards, making 0.5\% (6.5 products sold a day) pertaining to the noisy part of the distribution.}
\end{figure}

As we have seen in Fig. 6, the hazard rate of digital camera could be seen as stable during the whole data set.  The same assumption also holds for TV's and keyboards.  For these products, we assume that $h(t)$ is constant for all dates and estimate the constant hazard rate $\hat{h}$ by analyzing the lifetime distribution of products which died before January 1 2008 using the same method as described in Fig. 6.  For mouse products, we first divided the products into three subsets, 1) products which died before January 1 2006, 2) products which died during the period from January 1 2006 to December 31 2007 3) and the rest. We estimated the time varying hazard rates $\hat{h}_{1}$, $\hat{h}_{2}$ by estimating the lifetime distribution of the first two subsets 1) and 2) (see Fig. 8).  We then assumed $h(t)=\hat{h}_{1}$ for all $t< 1.1.2006$, $h(t)=\hat{h}_{2}$  for all $1.1.2006 \leq t < 1.1.2008$ and used these estimates to form our theoretical prediction
using Eq.~4.

We inserted these estimates into Eq. 4 and compared it with the empirical power-law exponents,
as shown in  Fig. 9.  In all cases, our theoretical prediction succeeds in capturing the increasing or stable trend in the time evolution of the power-law exponents.  The left panel of Fig. 8 shows the lifetime distribution for the mouse product and suggests that the hazard rate has dropped with time.  Note that putting $(\hat{r},\hat{\sigma},\hat{h}_{1})=(-0.09,0.4,0.049)$ in the formula would give $\mu=2.38$ and putting $(\hat{r},\hat{\sigma},\hat{h}_{2})=(-0.09,0.4,0.022)$ would give $\mu=2.25$ which is in good agreement with empirical values.  Hence, the decrease of the power-law exponent found for the mouse product could be attributed to the change in the hazard rate.  The analysis in this subsection shows that Malevergne et al.'s theory [30] is able to capture the variation of the power-law exponents with time and among product sets as well.

\begin{figure}[!h]
\begin{minipage}{0.4\hsize}
\begin{center}
\includegraphics[width=\hsize,bb= 12 4 152 127]{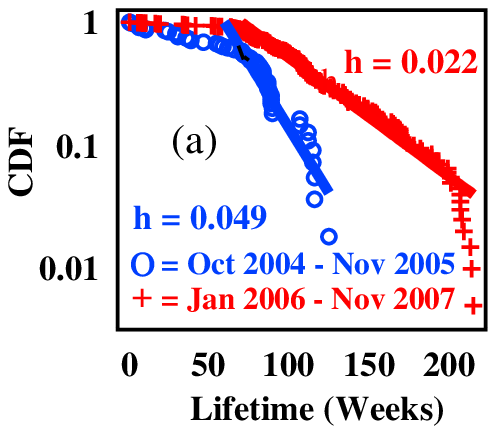}
\end{center}
%+22
\end{minipage}
\begin{minipage}{0.5\hsize}
\begin{center}
\includegraphics[width=\hsize,bb=15 18 319 219]{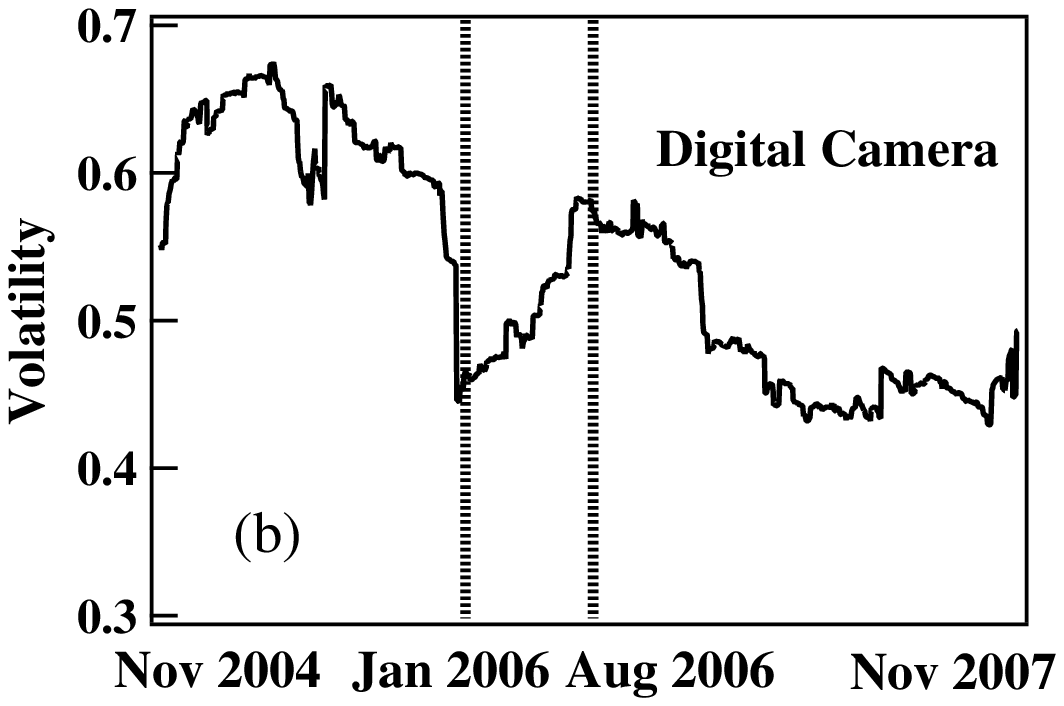}
\end{center}
%+14

\end{minipage}
\caption{(color online) The left panel (i.e. (a)) depicts the complementary cumulative distribution functions of lifetimes of the mouse product.  Blue circles denote the distribution of lifetimes of products that died during October 2004 to December 2005. The red crosses show  the distribution of lifetimes of products that died during January 2006 to December 2007. The straight lines show the maximum likelihood estimates, assuming an exponential distribution.  Only products which attained overall market share of 0.5\% or higher during their lifetimes were used when estimating the distribution of lifetimes, in order to focus on the top-selling products.  The right panel (i.e. (b)) depicts the time evolution of the volatility ($\sigma$) of market shares of digital cameras.  We can see an extraordinary drop in volatility during the time window January 16 2006 to August 8 2006.}
\end{figure}

\begin{figure}[!h]
\begin{minipage}{0.475\hsize}
\begin{center}
\includegraphics[width=\hsize,bb=  19 23 293 176]{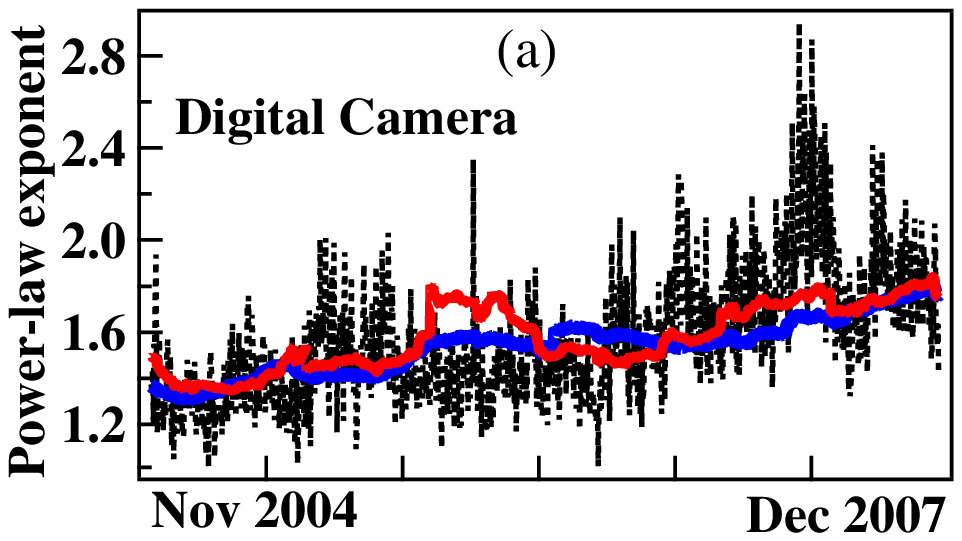}
%+0
\end{center}
%\caption{'ЂƂ'߂̂¸}
%\label{fig::oneichi}
\end{minipage}
\begin{minipage}{0.475\hsize}
\begin{center}
\includegraphics[width=\hsize,bb=  19 23 293 176]{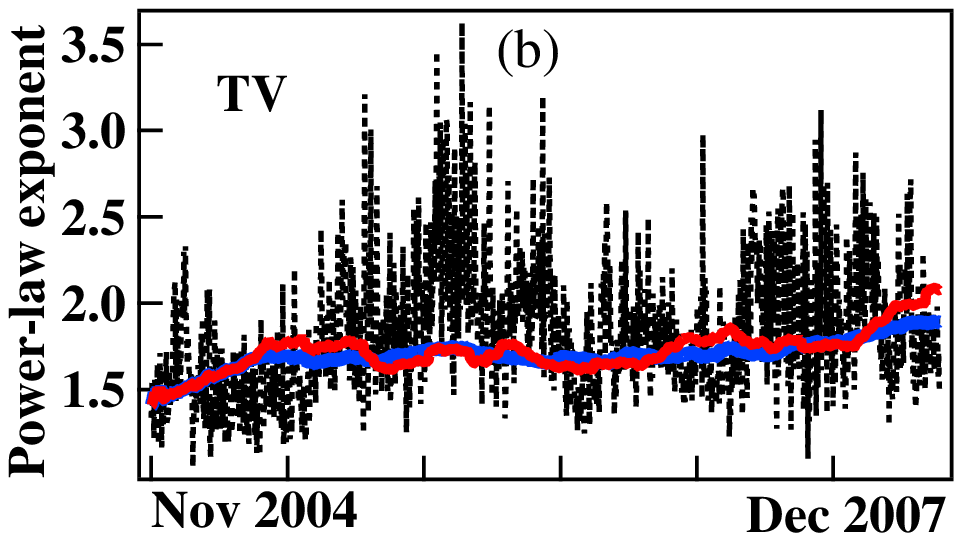}
%+30
\end{center}
%\caption{"ñ'–ڂ̐}}
%\label{fig:oneni}
\end{minipage}
\begin{minipage}{0.475\hsize}
\begin{center}
\includegraphics[width=\hsize,bb=  19 23 293 176]{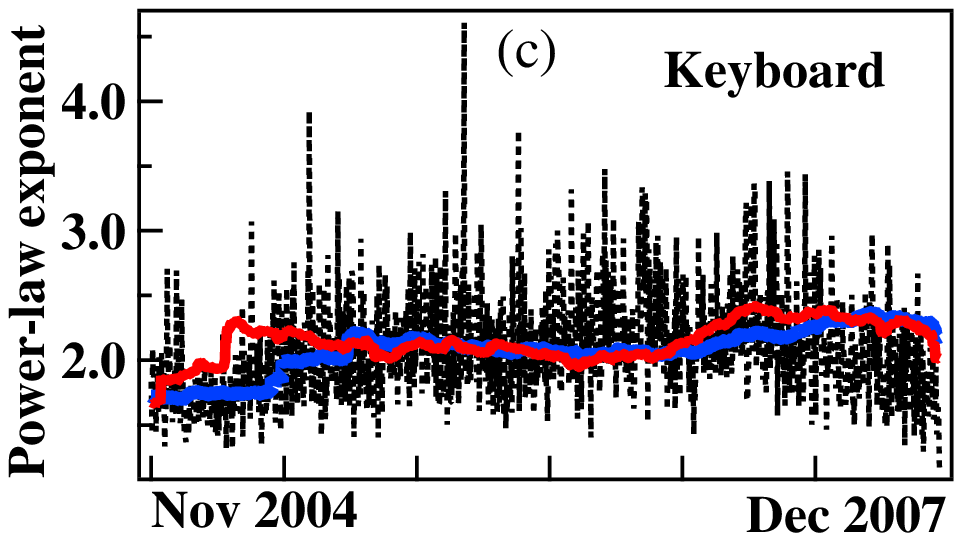}
\end{center}
%%\caption{'ЂƂ'߂̂¸}
%%\label{fig::oneichi}
\end{minipage}
\begin{minipage}{0.475\hsize}
\begin{center}
\includegraphics[width=\hsize,bb= 18 23 293 176]{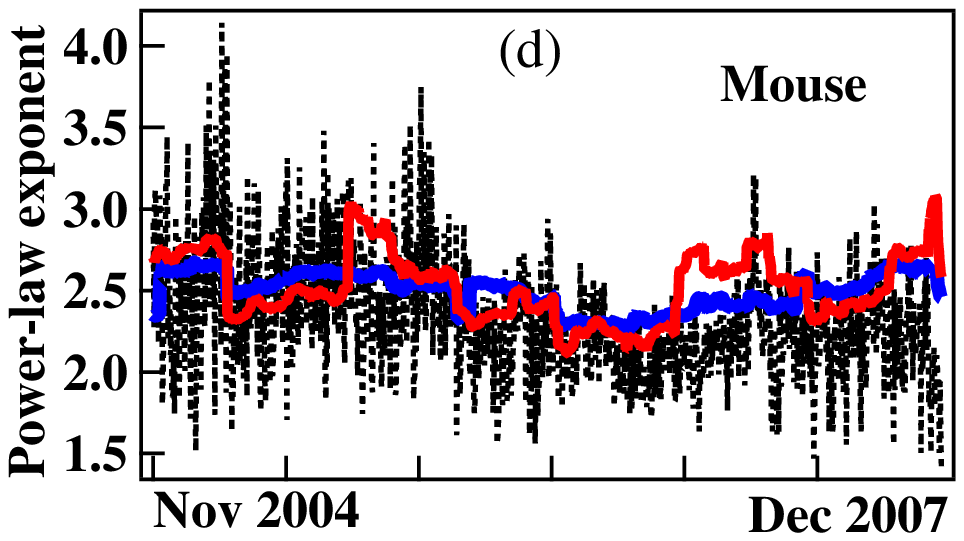}
\end{center}

\end{minipage}
\caption{(color online) Comparison between the maximum likelihood estimate $\mu(MLE)(t)$ and our theoretical prediction $\mu(TH)(t)$ for four time series, namely (a) digital camera, (b) TV, (c) keyboard and (d) mouse.  For each panel, the dashed line shows the estimated power-law exponent obtained from the empirical distribution.  The lower threshold for the validity of the power law tail of the distribution was estimated using the method described in Ref.~\cite{31}.  The red continuous line shows our theoretical exponent obtained for a running
window of 6 month duration. The blue line shows our theoretical exponent for a running time window of 1 year duration.}
\end{figure}

\subsection{Deviation from the pure power law form}

Hisano and Mizuno [31] analyzed the same data set and reported that the asymptotic behavior of the market share distribution of digital cameras during the period January 16, 2006 to August 8, 2006 deviates from the pure power-law form, but did not fully explain the cause of the deviation.  Notice that this period is exactly the same period when our theoretical power-law exponent $\mu(TH)$ calculated with a 6 month time window deviates from the empirical power-law exponent for digital camera, as can been seen in Fig. 9.  Equipped with the understanding of this paper, this deviation can now be understood by the failure of the delivery of new products, i.e., by a breakdown of the validity of assumption 3.

Indeed, top-selling products are usually forced to leave the market due to peer pressure (i.e. product competition) from newly born products which arises in a 6 month cycle (recall Fig. 4).  However during January 16, 2006 to August 8, 2006, many providers failed to deliver their new products, as we can see from the time evolution of birth flows (Fig. 4).  The lack of peer pressure could also be seen in the time evolution of the volatility ($\sigma$) of market shares as well (right panel of Fig. 8).  We can see that the volatility ($\sigma$)  in a 6 month time window indeed fell dramatically when there was a lack of new products,  and recovered afterward.  
The relationship between the lack of new products and the deviation from a pure power-law form can be 
rationalized by Malevergne et al.'s theory because an essential ingredient to generate a power-law distribution is indeed the birth flow of new products.  This observation shows that, not only the value of the power-law exponent, but  also the deviation from a genuine power-law form in the tail of the distribution provide us with informative insights about the corresponding system. The observed transient deviation from the pure power law form provides a crucial test
of the importance of the existence of a flow of births, that validates further the theory [30], in contrast with the alternative approaches [23,26,27].

\section{Concluding remarks}

Power-law distributions after their time of fame and fashion have been sometimes decried as too universal to really provide useful insights.  Here we have shown that by constructing a verifiable framework and decomposing the power-law exponent into its key ingredients, we were able to gain a richer understanding of how the empirical power-law distributions are formed and evolve with time in real world complex systems.    Our empirical analysis verifies the model proposed in [29,30] and opens the door for a wide range of applications in which power-law distributions are found in the presence of stochastic growth, birth and death.  

Our results also suggest that, rather than searching
for universality with a single power law exponent, the often observed variability of power law exponents could be used
as a diagnostic of the underlying mechanisms that generate the dynamics. By this, we mean the following.
If we have different realizations of a given system over time for instance that correspond to different exponents, then the values of these different exponents reveal the relative importance of birth, death and stochastic proportional growth, if indeed the generating mechanism of the system is based on these ingredients. By measuring directly on the system the proportional growth ingredients as in figure 2 and the distribution of lifetimes as in figures 5 and 7, we can verify the validity of the ingredients of the generating mechanism.  We can then check if the predicted exponents, based on the theoretical mechanism, match the observed ones. If yes, this cumulative evidence is a strong support for the concept that the exponents are deeply associated with birth, stochastic growth and death processes and their specific value indeed reveal, via the formula (4), the relative importance of the different ingredients. 
More work to explore the potential of this 
approach is of course needed.  

Given the generality of these ingredients, we believe that the prediction of the power law exponents provides new understandings of power law distributions, which will be insightful to many natural, economic and social systems.

\begin{acknowledgments}
The authors are grateful for Qunzhi Zhang and Ryan Woodard for helpful discussions and comments concerning this work.  This research is a part of a project entitled: Understanding Inflation Dynamics of the Japanese Economy, funded by JSPS Grant-in-Aid for Creative Scientific Research (18GS0101).  We would also like to thank BCN Inc. for providing its data.  Ryohei Hisano is supported by funding from the Japanese Student Services Organizations with a scholarship titled  ``Scholarship for Long-term Study Abroad 2010".
\end{acknowledgments}

% Our empirical analysis suggests that the variations of the exponents are associated with an increase of efficiency in the supply chain with time, which gives a rare insight into the inner organization of real life complex system/ecology tending to optimize to survive.

%\end{article}

\end{document}